\documentclass{article}

\usepackage[english]{babel}

\usepackage[letterpaper,top=2cm,bottom=2cm,left=3cm,right=3cm,marginparwidth=1.75cm]{geometry}

\usepackage{amsmath}
\usepackage{graphicx}
\usepackage[colorlinks=true, allcolors=blue]{hyperref}

\title{Virial clouds to explain rotational asymmetry in galactic halos}

\author{Asghar Qadir$^{1,*}$; Noraiz Tahir$^{1,\dagger}$; Francesco De Paolis$^{2,3,\ddagger}$; \\ Achille A. Nucita$^{2,3,\S}$\\
$^1$ School of Natural Sciences, National University of Science and Technology, H-12, \\ Islamabad, 46000, Pakistan\\
$^2$ Department of Mathematics and Physics ``E. De Giorgi'', University of Salento, \\ Via per Arnesano, I-73100  Lecce, Italy. \\
$^3$ INFN, Sezione di Lecce, Via per Arnesano, 73100 Lecce, Italy. \\
$^*$ asgharqadir46@gmail.com \\
$^\dagger$ noraiztahir78637@gmail.com \\
$^\ddagger$ francesco.depaolis@le.infn.it \\
$^\S$ achille.nucita@le.infn.it \\
}

\begin{document}
\maketitle

\begin{abstract}
The rotation of the galactic objects has been seen by asymmetric Doppler shift in the CMB data. Molecular hydrogen clouds at virial temperature may contribute to the galactic halo dark matter and they might be the reason for the observed rotational asymmetry in the galactic halos. We present a method to constrain the parameters of these virial clouds given that they are composed of a single fluid. The method is such that it should be possible to extend it to more than one fluid.
\end{abstract}

\section{Introduction}
Half of the baryonic matter in the Universe is dark \cite{1,2,3,4}, some of it may consist of molecular hydrogen clouds left over from the process of star formation, which would continue to collapse till their virial temperature reaches the CMB. To observe them one could look for $\gamma$-rays scintillation due to strikes by cosmic rays on ${\rm H_{2}}$ molecules. A  $\gamma$-ray corona around the Milky Way \textit{was} discovered \cite{5} with the expected angular distribution \cite{6}, but it was not not clear how much, if any, of the carpet might be due to the proposed clouds. An alternative suggestion was to look for an asymmetric Doppler shift of these clouds in M31 \cite{7}.

The predicted effect was seen at a sufficient but low level in the WMAP data in 2011 \cite{8} and confirmed in the Planck data in 2014 \cite{9}. Objections that the blue shift may be due to ``hot spots" were countered by the argument that the red shift could not be ``cold spots" and were soon dispelled. The precision was sufficient to be able to use for studying the rotation of the M31. This was followed by the same effect being seen in NGC 5128, M33, M82, M81 \cite{10,11,12,13}. However, the verification of the effect that it is due to the molecular clouds which led to its prediction, does not prove that it is due to the cause that led to the  prediction. It could be partially or totally due to other interstellar matter. To be able to determine the nature of the cause, we need to model the molecular hydrogen clouds or dust clouds produced by the mechanism mentioned, or some mixture of the two, and compare with the data. One could then try to look at whether contamination of the spectrum by radiation from higher temperature sources could be distinguished.

\section{The Virial Cloud Model}
As these clouds are immersed in a heat bath of the CMB, they are isothermal. Since they have formed due to the Jeans instability, they must have a precisely defined Jeans radius and consequent Jeans mass. The density profile would then be given by the solution of the Lane-Emden equation subject to the boundary conditions that the density be literally {\it zero} (not approximately zero) at the boundary and the density profile be flat at the core. The latter condition says that the derivative of the density be zero at the core. We call them ``virial clouds''. Since James Jeans had applied the analysis for the formation of a normal star, he had taken the boundary condition that the density of the cloud merge with that of the interstellar medium (ISM) (which has been called ``the Jeans fiddle'') \cite{14}. In that case the central density had to be put in on an ad hoc basis. We do not need to do this. Further, we will want to extend the analysis (in later work) to a two-fluid model. As such, we obtain the equations from first principles in any case.

\begin{figure}[ht]
	\includegraphics[width=11cm]{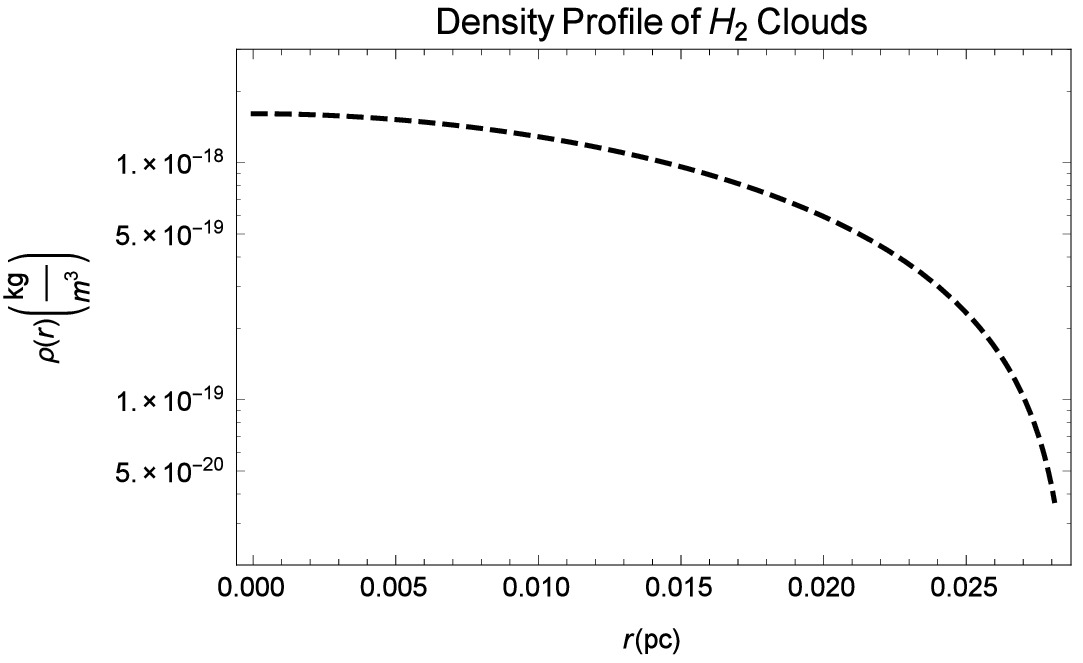}
	\caption{The density of the virial molecular hydrogen cloud. It decreases monotonically from the central density, ${\displaystyle \rho_{c} \simeq 1.60 \times 10^{-18}~ kgm^{-3}}$, to zero at $r=R_J\simeq0.030~{\rm pc}$.
		\label{f1}}
\end{figure}

For an isothermal gas sphere, the temperature, $T$, is directly related to the mass of the molecule, $m_H$ in our case, and the average speed. The speed of the perturbation would have to be taken to be the speed of sound in the cloud, $c_s$. This gives
\begin{equation}
c_s=\sqrt{\frac{\gamma kT_{CMB}}{m_H}}~, 	
\label{e1}
\end{equation}
where $\gamma$ is the adiabatic factor given by the ratio of the heat capacity at constant pressure to that at constant volume. At the densities and temperatures that we are considering no higher degrees of freedom would
be excited and we would only have the translational degree, making the ideal gas approximation extremely good and so $\gamma=5/3$. For our purpose, $m_H\approx2.016~{\rm g/mol}$ and $T\approx2.726^oK$, so
\begin{equation}
c_{s}\simeq 1.110 \times 10^{4}~{\rm km~s^{-1}}~. 	
\label{e2}
\end{equation}
Now the virial theorem $2K+U=0$ gives the Jeans mass as \cite{15}
\begin{equation}
M_J^2\simeq \left(\frac{81}{32\pi\rho_c}\right)\left(\frac{3c_s^2}{5G}\right)^3~,
\label{e3}
\end{equation}
and the corresponding Jeans radius as
\begin{equation}
R_J^2=\frac{27c_s^2}{20\pi\rho_cG}~,
\label{e4}
\end{equation}
where, $\rho_c$ is the central density of the virial cloud, and $G$ is Newton's gravitational constant. We need to use the canonical ensemble distribution in order to calculate the density profile of the clouds which is given by the relation \cite{16}
\begin{equation}
f(r,p)=\frac{1}{h^{3N}N!}\frac{1}{Z}exp\left(-\frac{H(r,p)}{kT_{CMB}}\right)~,
\label{eq5}
\end{equation}
where, $H(r,p)$ is the Hamiltonian of the system. The partition function $Z$ is the normalization factor, thus being the integral over phase space of the rest of this distribution function. The Hamiltonian is position dependent
\begin{equation}
H(r,p)=\frac{p^2}{2m_H}+\frac{GM(r)m_H}{r}~,
\label{eq6}
\end{equation}
and so the density distribution, is given by
\begin{align}
\rho(r)=8m_H^{5/2}\left(\frac{G~\rho_c}{3k~T_{CMB}}\right)^{3/2} exp\left(-\frac{GM(r)m_H}{rkT_{CMB}}\right),	
\label{eq7}
\end{align}
where $M(r)$ is the mass of the cloud interior to $r$ given by
\begin{equation}
GM(r)=4\pi\int_0^r\rho(q)q^2dq~.
\label{eq8}
\end{equation}
Our problem, then, is that the density is defined in terms of the integral of the density, i.e. we have an integral equation. Taking the natural logarithm of eq.(7) we obtain
\begin{equation}
\int_0^r q^2\rho(q)dq=-\frac{rkT_{CMB}}{4\pi Gm_{H}}\ln
\left(\frac{\rho_(r)}{\zeta}\right)~,
\label{eq9}
\end{equation}
where
$\zeta=8m_H^{5/2}(G\rho_c/3kT_{CMB})^{3/2}$. Taking the derivative of the above equation with respect to $r$, and substituting eq.(7) we obtain the ordinary differential equation
\begin{equation}
r\frac{d\rho(r)}{dr}-r^2\left(\frac{4\pi Gm_{H}}{kT_{CMB}}\right)\rho^2(r)
-\rho(r)\ln\left(\frac{\rho(r)}{\zeta}\right)=0~.
\label{eq10}
\end{equation}
It will be noticed that this is a nonlinear differential equation that cannot be solved by separation of variables. However, we do have the initial condition  $\rho^{\prime}(r)|_{r=0}=0$ and the boundary condition that $\rho(R_J)=0$. As such, we can solve it numerically by assuming a central density and the initial condition. If the density at the boundary is too high, we lower the central value and try again, till it becomes too low. At that stage we raise the central density a bit till it again becomes too high. Iteratively we bring the boundary density close enough to zero to be acceptable.  The resulting density profile is shown in Fig.\ref{f1}. One sees that the central density of the cloud is ${\displaystyle \rho_{c} \simeq 1.60 \times 10^{-18}~ kgm^{-3}}$, the radius is $R_J\simeq0.030{\rm pc}$, and the mass is $M_J\simeq0.798M_{\odot}$.

\section{Luminosity of the Clouds}
Since the virial clouds are at the CMB temperature, their luminosity is the CMB luminosity, except for their Doppler shift. Thus, we only need to look at the differential frequency shift, $\Delta\nu/\nu=-(v_{rot}/c)\cos\theta$, where $v_{rot}$ is the rotational velocity of the galactic halo, $c$ the speed of light and $\theta$ the angle between the direction of the velocity and the line of sight. Now $\Delta\nu/\nu=\Delta E/E=\Delta T/T$, so the differential energy is given by $\Delta E=(\Delta T/T)E$, where $E$ is the energy in the radiation for a patch of the sky corresponding to the size of the cloud. Now, as the luminosity, $L$, is proportional to the energy, we get the luminosity of the CMB for that patch. Then the luminosity of the cloud will be \cite{17}
\begin{equation}
L_c=(\Delta\nu/\nu)L\simeq1.486\times10^{-4}L\cos\theta~.
\label{eq11}
\end{equation}
Thus, as we scan across the halo, the luminosity will not change linearly but with $\cos\theta$. We can use this check for the Doppler shift explanation and obtain the energy associated with each cloud. This would directly give an estimate of the total number of clouds being seen.

\section{Results and Discussions}
The virial clouds are completely constrained by physical requirements, and so precisely determinable. They provide a satisfactory explanation of the observed asymmetry of the CMB Doppler shift in galactic halos. We have only considered pure molecular hydrogen clouds, but we also need to model these molecular hydrogen clouds contaminated with more or less interstellar dust which will be dealt with elsewhere. We have not discussed the possible contamination of radiation by matter at high temperature. That needs to be done. The effect is probably negligible, but cannot be neglected without checking. One also needs to investigate how the clouds would evolve over billions of years, from the time when the CMB temperature was higher than it is now to the present value. Precisely how did the clouds lose energy over this period?

\section*{Acknowledgments}
We are grateful to the Salento University of Lecce for hospitality at Lecce, where part of this work was done, and AQ is grateful to IUPAP, ESA, ICRANet and AS-ICTP for support to participate in MG15.
%\begin{thebibliography}{000} %for 3 digits

\end{document}